\documentclass[preprint,showpacs,preprintnumbers,amsmath,amssymb]{revtex4}

\usepackage{graphicx}
\usepackage{dcolumn}
\usepackage{bm}

\begin{document}


\title{Peripheral fragmentation of the $^8$B nuclei fragmentation 
at an energy of 1.2~A~GeV in nuclear track emulsion}

\author{R.~Stanoeva}
   \email{stanoeva@lhe.jinr.ru}    
     \homepage{http://becquerel.jinr.ru}
     \homepage{http://pavel.jinr.ru}
  \affiliation{Joint Insitute for Nuclear Research, Dubna, Russia}
 \author{V.~Bradnova}
   \affiliation{Joint Insitute for Nuclear Research, Dubna, Russia}
\author{S.~Vok\'al}
   \affiliation{P. J. \u Saf\u arik University, Ko\u sice, Slovak Republic} 
 \author{P.~I.~Zarubin}
\affiliation{Joint Insitute for Nuclear Research, Dubna, Russia} 
 \author{I.~G.~Zarubina}
   \affiliation{Joint Insitute for Nuclear Research, Dubna, Russia}    
\author{N.~A.~Kachalova}
   \affiliation{Joint Insitute for Nuclear Research, Dubna, Russia} 
\author{A.~D.~Kovalenko}
   \affiliation{Joint Insitute for Nuclear Research, Dubna, Russia}  
 \author{A.~I.~Malakhov}
   \affiliation{Joint Insitute for Nuclear Research, Dubna, Russia} 
\author{G.~I.~Orlova}
   \affiliation{Lebedev Institute of Physics, Russian Academy of Sciences, Moscow, Russia} 
\author{N.~G.~Peresadko}
   \affiliation{Lebedev Institute of Physics, Russian Academy of Sciences, Moscow, Russia} 
\author{P.~A.~Rukoyatkin}
   \affiliation{Joint Insitute for Nuclear Research, Dubna, Russia} 
\author{V.~V.~Rusakova}
   \affiliation{Joint Insitute for Nuclear Research, Dubna, Russia}
\author{E.~Stan} 
   \affiliation{Institute of Space Sciences, Magurele, Romania}
\author{M.~Haiduc}
   \affiliation{Institute of Space Sciences, Magurele, Romania}
 \author{I.~Tsakov}
   \affiliation{Institute for Nuclear Research and Nuclear Energy, Sofia, Bulgaria}  
\author{S.~P.~Kharlamov}
   \affiliation{Lebedev Institute of Physics, Russian Academy of Sciences, Moscow, Russia} 

\date{\today}

\begin{abstract}
\indent 
The results of investigations dealing with the charge topology of the fragments produced 
in peripheral dissociation of relativistic $^8$B nuclei in emulsion are presented. 
55 events of peripheral  dissociation of the $^8$B nucleus were selected from the events 
which do not involve the production of the target-nucleus fragments and mesons (\lq\lq white\rq\rq ~stars). 
A leading contribution of the  $^8$B$\rightarrow^7$Be+p mode having the lowest energy threshold
was revealed on the basis of those events. Information on the relative probability of dissociation 
modes with a larger multiplicity was obtained. The dissociation of a $^7$Be core in $^8$B
indicates an analogy with that of the free $^7$Be nucleus.\par

\indent The transverse momentum distributions of  the fragments from the $^8$B$\rightarrow^7$Be+p 
dissociation mode were obtained. Their small average value, $<$P$_t>$= 52 MeV/c, 
in the c.m.s. suggests a low binding energy of the external proton in the $^8$B nucleus.
 An indication for a strong azimuthal correlation of the $^7$Be and p fragments was got.
\par
\end{abstract}

\maketitle
\section{\label{sec:level1}Introduction}
\indent A peculiar feature of the $^8$B nucleus is known to consist in a low proton separation energy. 
This fact suggests that in $^8$B  there can possibly exist a proton halo which is spacially separated 
 from a core represented by the $^7$Be nucleus. In relativistic fragmentation processes $^8$B$\rightarrow^7$Be
 such a loose bound  results in very narrow momentum distribution of $^7$Be nuclei \cite{Schwab95,Smedberg99}. 
In just the same way as in the case of exotic nuclei, a possible increase in the $^8$B nucleus radius 
caused by the proton halo could have lead to a relative growth  in its interaction cross section. 
However this increase is not observed experimentally. Therefore a more detailed study of the $^8$B nucleus, 
namely its cluster structure, is especially urgent. For example, a proton halo  can arise on the basis 
of the deuteron cluster from an odd neutron with an external proton. The $^7$Be and $^8$B spins equal 
to 3/2 and 2, respectively, point to this possibility.  
\par
\indent The nature of the proton halo can be clarified by means of a more thorough analysis of
the  $^8$B relativistic fragmentation in peripheral interactions. The interactions of this type are provoked 
 either in electromagnetic and diffraction processes, or in nucleon collisions  at small overlapping of 
the colliding nucleus densities. A fragmenting nucleus gains an excitation spectrum near the cluster 
dissociation thresholds. In the kinetic region of fragmentation of a relativistic nucleus there are produced
nuclear fragment systems the total charge of which  is equal to the parent-nucleus charge. 
A relative intensity of formation of fragments of various configurations makes it possible to estimate the 
importance of different cluster modes. For the $^8$B nucleus the following modes $^7$Be+$^1$H, 
$^{4,3}$He+$^{3}$He+$^{1,2}$H, $^{6}$Li+$^{2,1}$H, $^{4,3}$He+$^{2,1}$H+$^{2,1}$H, and 5$^{2,1}$H. are possible.
\par
\indent The opening angle of the relativistic fragmentation cone is determined by the Fermi-momenta of the nucleon 
clusters in a nucleus. Being normalized to the mass numbers they are concentrated  with a few percent dispersion 
near the normalized momentum of the primary nucleus. When selecting events with dissociation of a projectile into
 a narrow fragmentation cone we see that target-nucleus non-relativistic fragments either are absent 
 (\lq\lq white\rq\rq ~stars in \cite{Adamovich05}), or their number is insignificant. The target fragments are 
easily separated from the fragments of a relativistic projectile since their fraction in the angular 
relativistic fragmentation cone is small and they possess non-relativistic momentum values.
\par
\indent In the peripheral fragmentation of a relativistic nucleus with charge Z the ionization induced by
 the fragments can decrease down to a factor Z, while the ionization per one track  -  down to Z$^2$. 
Therefore experiment should provide an adequate detection range. In order to reconstruct an event,
 a complete kinematic information about the particles in the relativistic fragmentation cone is needed which,
 e.g., allows one to calculate the invariant mass of the system. The accuracy of its estimation decisively 
depends on the exactness of the track angular resolution. To ensure the best angular resolution, it is necessary 
that the detection of relativistic fragments  should be performed with the best spacial resolution.
\par
\indent The nuclear photoemulsion method, which underlies the BECQUEREL project at the JINR nuclotron \cite{web},
 well satisfies the above-mentioned requirements. It is aimed at a systematic search for peripheral fragmentation modes
 with statistical provision at a level of dozens events, their classification and angular metrology. 
Emulsions provide a record special resolution (about  0.5 $\mu$m) which allows one to separate the charged 
particle tracks in the three-dimensional image of an event within one-layer thickness (600 $\mu$m) and 
ensure a high accuracy of angle measurements. The tracks of relativistic H and He nuclei are separated by sight. 
As a rule, in the peripheral fragmentation of a light nucleus its charge can be determined by the sum of 
the charges of relativistic fragments. Multiple-particle scattering measurements on the light fragment tracks 
 enable one to separate the H and He isotopes. The analysis of the products of the relativistic fragmentation 
of neutron-deficient isotopes has some additional advantages owing to a larger fraction of observable nucleons and 
minimal Coulomb distortions. Irradiation details and a special analysis of interactions in the BR-2 emulsion
are presented in \cite{Adamovich99,Adamovich04}. In what follows, we give the first results of the study of the $^8$B
nucleus fragmentation at a 1.2A GeV energy which are obtained with the use of a part of the material analyzed. 
In particular, we found the final charged state and fragment emission angle distributions. A comparison with the 
data for $^{10}$B \cite{Adamovich05,Adamovich04} and $^8$B \cite{Perasadko06} nuclei was made.
\par
\begin{figure}
\includegraphics[width=160mm]{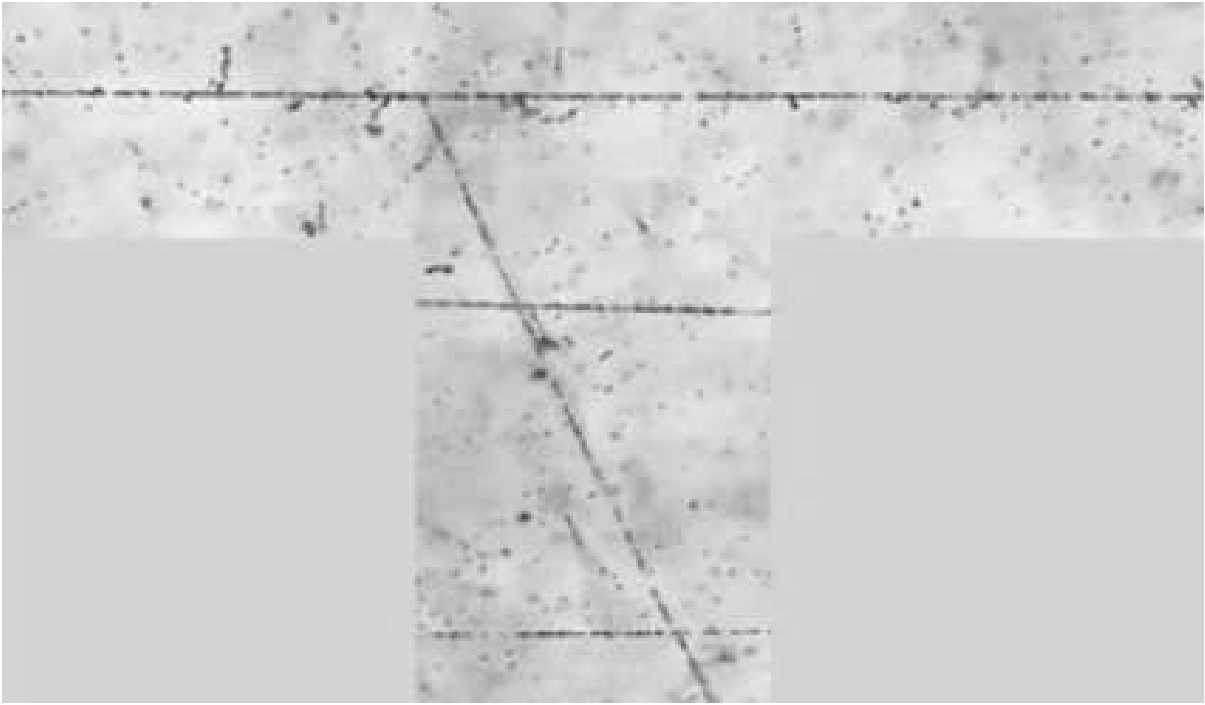} 
\caption{\label{fig:1} Micro-photograph of the $^{10}$B  fragmentation at an energy of 1.2 A GeV 
on a $^{1}$H  nucleus in emulsion which was made by means of the  PAVIKOM (FIAN) complex. The primary $^{10}$B 
nucleus track does not  practically change its direction after the interaction point. The recoil proton track 
 has a large emission angle. The process is identified as $^{10}$B+p$\rightarrow^8$B+2n+p.}

\end{figure}
\section{\label{sec:level2}Exposure of emulsion to a $^8$B beam}

\indent In investigations with the use of emulsion exposed to secondary beams it is necessary that the 
identification of the nuclei being studied should be rather simple. For this purpose, to form a $^8$B  beam at 
the JINR Nuclotron, use was made of the fragmentation process of a primary $^{10}$B  nucleus beam of an 
 energy of 1.2A GeV  on a polyethylene target. Such processes were observed in the study of the $^{10}$B
interactions in emulsion \cite{Adamovich04}. As an example,  Fig.~\ref{fig:1} displays a microphotograph  of an event 
of the primary $^{10}$B nucleus fragmentation into a nucleus with a charge of  5 which entails the recoil proton 
track  As far as , on the one hand, a $^{9}$B isotope in the bound state does not exists and, on the other hand, 
it is necessary to compensate the recoil proton transverse momentum  the  most probable interpretation of 
this event is to suppose that two neutrons should escape from $^{10}$B and a $^{8}$B should be formed. 
\par
\indent The absence  of the $^{9}$B isotope among relativistic fragments turns out to be useful when separating 
the $^{8}$B beam from the $^{10}$B beam by the magnetic rigidity (the difference is about 20\%). The beam channel 
at the JINR nuclotron which was used for irradiation was expected to have a suitable momentum acceptance  
of about 2-3\%. This suggestion was confirmed when setting the channel to $^{9}$B separation: the boron nuclei 
practically vanished. But when setting to $^{10}$B and $^{8}$B separation they appeared anew. An analysis of 
irradiated emulsions gave an additional proof of this suggestion  that was the absence of 
\lq\lq white\rq\rq ~stars with the topology of the relativistic fragments of the H+He and He+Li pairs originating 
from possible $^{6}$Li and $^{10}$B admixtures. As a scintillation monitor showed the contribution from the $^{7}$B 
nuclei which have a close magnetic rigidity (the difference is about  10\%) to the formed $^{10}$B beam was less than 10\%.
Their contribution to the number of \lq\lq white\rq\rq ~stars was excluded due to the charge topology. 
The most intense background from the beam tracks presented by the $^{3}$He nuclei was rejected in scanning emulsion by sight.
\par

\section{\label{sec:level3}The charge topology of $^{8}$B dissociation}
\indent Over the viewed-track length of 123 m we found 929 events of inelastic interactions. The mean free path was 
estimated to be $\lambda$=13.3$\pm$0.4 cm. The latter value correspond to the inelastic cross section for a $^{8}$B
 nucleus with a normal density. The method of viewing over the primary-particle track aimed at the search for interactions 
in emulsion shows that the frequency of emergence of the events with a given charge topology defines the probability 
of various dissociation channels. Among the found events there are 320 stars in which the total charge of the relativistic
fragments in a 8$^{\circ}$ fragmentation cone Q satisfies the condition Q $>$ 3. These stars were attributed to the number 
of peripheral dissociation events N$_{pf}$.  The  N$_{pf}$ relativistic fragment distribution of over charges N$_Z$
is given in Table \ref{tab:1}. There are given the data  for 256 events containing the target-nucleus fragments  - N$_{tf}$,
as well as for 64 events which contain no target-nucleus fragments (\lq\lq white\rq\rq ~stars )-- N$_{pf}$. The role of the 
channels with multiple relativistic fragments  N$_{Z}>$2 is revealed to be dominant for the N\lq\lq white\rq\rq ~stars. 
Of peripheral events, the \lq\lq white\rq\rq ~stars N$_{ws}$ (Table \ref{tab:1}) are of very particular interest. They are not 
accompanied  by the target-nucleus fragment tracks and makes it possible to clarify the role of different cluster degrees
of freedom at a minimal excitation of the nuclear structure.\par
\indent Two \lq\lq white\rq\rq ~stars with Q=4  may be referred to the background $^{7}$Be nucleus contribution (Table \ref{tab:1}).
An analysis of the emulsions exposed to a $^{7}$Be beam revealed [7] that the probability of peripheral $^{7}$Be
interactions involving target-fragment production is about half as much  as  that involving no target-fragment production.
This means that the main contribution to the number  N$_{tf}$  with Q=4 must be given by the $^{8}$B interactions which 
occurred in the fragmentation cone with decreasing charge by unity. 
\par
\indent The main dissociation branch are the events with Q=5. The key difference between N$_{tf}$ and N$_{ws}$ for this
 group of events is revealed in a two-particle mode Z=4+1. The latter is unambiguously interpreted as $^8$B$\rightarrow^7$Be+p
 because of neutron deficiency. Its fraction sharply increases when selecting  \lq\lq white\rq\rq ~stars: from 10\%
 for the case of the presence of target fragments up to 44\%  for \lq\lq white\rq\rq ~stars. The threshold of this mode is the lowest
 one which is seen from the fact that it  dominates  the most peripheral events. 
\par
\indent The distribution of the probability of the \lq\lq white\rq\rq ~stars production by $^8$B nuclei can be compared
 with similar data for the $^{10}$B nuclei of an energy of  1A GeV \cite{Adamovich05,Adamovich04}. The fraction of the 3-prong 
mode  $^{10}$B$\rightarrow$2He+$^{1,2}$H amounts to 73\% for a 40\% deuteron contribution. The probability of the 
2-prong mode $^{10}$B$\rightarrow^9$Be+$^{1}$H  was found to be as small as 2\%. The difference is due to a lower binding
 energy of the deuteron as compared to that of the proton in the $^{10}$B nucleus. On the contrary, for the $^{8}$B 
isotope there is a large yield of the two-body mode $^8$B$\rightarrow^7$Be+p because the binding energy of the external 
proton is low enough. In such a way, the  $^8$B nucleus manifests its structure through a more intense formation  of 
\lq\lq white\rq\rq ~stars with N$_Z$=2 .Among the Q=5 and N$_Z$=2 events, one event among N$_{tf}$,
 namely Li+He, may not be attributed to the $^8$B interaction.     
\par
\indent The probabilities of  the \lq\lq white\rq\rq ~stars fragmenting via the 2He+H and He+3H channels for the $^8$B
 nucleus were turned out to be 22\% and 25\%, respectively, while for the $^{10}$B nucleus  the probabilities of the same
 modes  were 73\% and 12\%. Such a noticeable difference is due to the presence  of a $^{3}$He cluster in $^{8}$B
  whose dissociation threshold is essentially lower than that for $^{4}$He.
\par

\begin{table}
\caption{\label{tab:1} The charge topology distribution of the number of  interactions of the peripheral 
N$_{pf}$ type (N$_{pf}$=N$_{tf}$+N$_{ws}$), which were detected in an emulsion exposed to a second $^8$B nucleus beam. Here Q is the total
charge of relativistic fragments in a 8$^{\circ}$ angular cone in an event, N$_{Z}$ the number of fragments with charge Z
 in an event, N$_{ws}$ the number of "white" stars, N$_{tf}$ the number of events involving the target fragments, 
N$_{ws}$ the number of \lq\lq white\rq\rq ~stars.}

\begin{tabular}{c|c|c|c|c|c|c|c}
\hline\noalign{\smallskip}
\hline\noalign{\smallskip}

~~Q	~~&~~N$_{5}$~~&~~N$_{4}$~~&~~N$_{3}$~~&~~N$_{2}$~~&~~N$_{1}$~~&~~N$_{tf}$~~&~~N$_{ws}$\\
~~7	~~&~~-	~~&~~-	~~&~~-	~~&~~1	~~&~~5	~~&~~1  ~~&~~-\\
~~6	~~&~~-	~~&~~-	~~&~~-	~~&~~2	~~&~~2	~~&~~8	~~&~~2\\
~~6	~~&~~-	~~&~~-	~~&~~-	~~&~~1	~~&~~4	~~&~~6	~~&~~4\\
~~6	~~&~~-	~~&~~-	~~&~~-	~~&~~-	~~&~~6	~~&~~1	~~&~~-\\
~~5	~~&~~-	~~&~~-	~~&~~-	~~&~~1	~~&~~3	~~&~~61	~~&~~14\\
~~5	~~&~~-	~~&~~-	~~&~~-	~~&~~2	~~&~~1	~~&~~44	~~&~~12\\
~~5	~~&~~-	~~&~~-	~~&~~1	~~&~~-	~~&~~2	~~&~~8	~~&~~-\\
~~5	~~&~~-	~~&~~-	~~&~~1	~~&~~1	~~&~~-	~~&~~1	~~&~~-\\
~~5	~~&~~-	~~&~~1	~~&~~-	~~&~~-	~~&~~1	~~&~~17	~~&~~24\\
~~5	~~&~~1	~~&~~-	~~&~~-	~~&~~-	~~&~~-	~~&~~17	~~&~~1\\
~~5	~~&~~-	~~&~~-	~~&~~-	~~&~~-	~~&~~5	~~&~~21	~~&~~4\\
~~4	~~&~~-	~~&~~-	~~&~~-	~~&~~-	~~&~~4	~~&~~5	~~&~~1\\
~~4	~~&~~-	~~&~~-	~~&~~-	~~&~~2	~~&~~-	~~&~~24	~~&~~1\\
~~4	~~&~~-	~~&~~-	~~&~~-	~~&~~1	~~&~~2	~~&~~42	~~&~~-\\

\hline\noalign{\smallskip}
\hline\noalign{\smallskip}
\end{tabular}
\end{table}

\begin{table}
\caption{\label{tab:2}The charged dissociation mode distribution of the \lq\lq white\rq\rq ~stars
 produced by the $^7$Be and $^8$B nuclei. To make the comparison more convenient, for the $^8$B nucleus
 one H nucleus is eliminated from the charged mode and  the  channel fractions are indicated.}

\begin{tabular}{c|c|c|c|c}
\hline\noalign{\smallskip}
\hline\noalign{\smallskip}
	~~Z	~~      & ~~$^7$Be~~& ~~\% ~~ & ~~$^8$B (+H)~~ & ~~\% ~~ \\
	~~2He~~	    & ~~41~~	& ~~43~~  & ~~12~~         & ~~40~~ \\
	~~He+2H	~~  & ~~42~~    & ~~45~~  & ~~14~~         & ~~47~~ \\
	~~4H~~	    & ~~2~~	    & ~~2~~	  & ~~4~~	       & ~~13~~ \\
	~~Li+H~~	& ~~9~~	    & ~~10~~  & ~~0~~		   & ~~0~~ \\
\hline\noalign{\smallskip}
\hline\noalign{\smallskip}
\end{tabular}
\end{table}

\indent The presence of the \lq\lq white\rq\rq ~stars with more than  two N$_Z>$2 fragments may be explained by the $^{7}$Be
 core dissociation.  In order to verify this suggestion, Table \ref{tab:2} gives the  relativistic fragment charge distribution in 
the \lq\lq white\rq\rq ~stars for $^{7}$Be  \cite{Adamovich05,Perasadko06} and $^{8}$B nuclei. The $^{8}$B events are presented without 
one single-charged relativistic fragment, that is a supposed proton halo. The identical fraction of the two main 2He and 
He+2H dissociation channels is observed for $^{7}$Be and $^{8}$B nuclei which points out that the $^{8}$Be core excitation 
is independent of the presence of an additional loosely bound proton in the $^{8}$B nucleus.
\par
\indent The observation of 4 unusual  Q=5 \lq\lq white\rq\rq ~stars with a total $^{8}$B$\rightarrow$5H disintegration 
 is noteworthy (Table \ref{tab:1}). This process leads to a breakup of  two He clusters and has a high energy threshold. Earlier,
 the events of such a type were already observed  for  $^{7}$Be$\rightarrow$4H and $^{10}$B$\rightarrow$5H \cite{Adamovich04}. 
The neutron deficiency in the $^{8}$B  nucleus makes the observability of nucleons far better. In addition,  the Coulomb 
repulsion in the fragment system becomes stronger. The formation of such H nucleus ensembles can be the basis of the 
multiple fragmentation of heavier nuclei with a large number of fragments. 
\par 
\indent The production of \lq\lq white\rq\rq ~stars  with Q=6  (Table \ref{tab:1}) may be due to a $^{10}$C admixture in the composition
 of the beam used. The $^{10}$C nuclei could be produced  through a $^{10}$B$\rightarrow^{10}$C charge exchange in
 the target intended for the $^{8}$B generation and could be captured to a second beam because of a small difference
 in the magnetic rigidity as compared with $^{8}$B (about 4\%)  and their proper momentum dispersion. 
The Q=6 \lq\lq white\rq\rq ~stars contain no fragments with Z$>$2. Their topology does not contradict the assumption about 
the dissociation of a carbon isotope with a $^{8}$B core ($^{10}$C$\rightarrow^{8}$Be+2p) or with the disintegration of 
one of the He clusters ($^{10}$C$\rightarrow^{4}$He+2H+2p). This fact points to a possible formation of a $^{10}$C 
beam ($^{10}$B$\rightarrow^{10}$C) under the conditions which are convenient  for performing investigations in emulsion. 
A  Q=7 "white" star may be referred to the production  of a charged meson pair. 
\par

\section{\label{sec:level4}Angular characteristics of $^7$Be+p}
\indent The \lq\lq white\rq\rq ~stars of  $^8$B$\rightarrow^7$Be+p (Table \ref{tab:1}) give a pair of observed tracks with a small 
angular deviation with respect to the primary nucleus track. Measurements gave the mean values of the polar emission 
angles $<\theta_p>$=2.0$^{\circ}$ for protons and $<\theta_{Be}>$=0.4$^{\circ}$ for $^{7}$Be nuclei.
\par

\begin{figure}
\includegraphics[width=160mm]{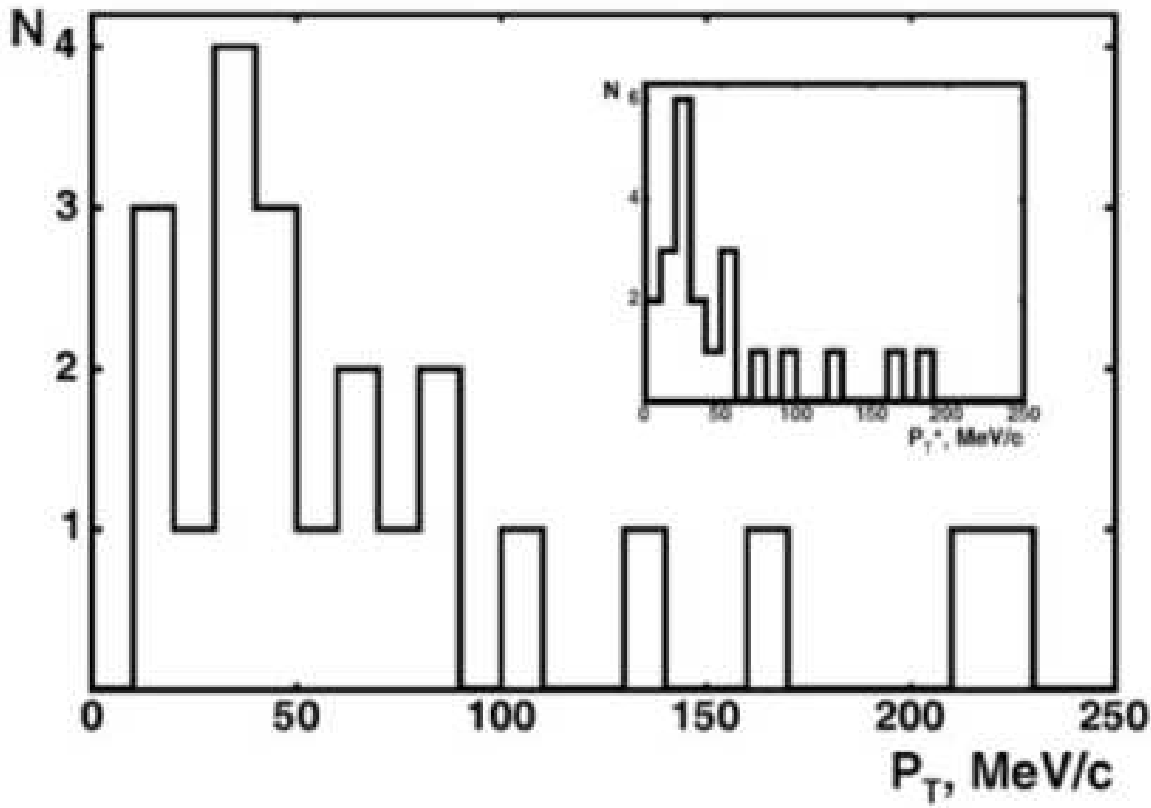}  
\caption{\label{fig:2} The P$_t$ transverse momentum distribution of the protons produced in \lq\lq white\rq\rq ~stars 
$^8$B$\rightarrow^7$Be+p. In the insertion the same  P$_t^*$ distribution is given in the c.m.s. of $^7$Be+p.}
\end{figure}

\indent The angular measurements permit one to reconstruct  with a good accuracy the spectra of transverse momenta 
P$_T$ by the equation P$_T$=AP$_0$sin$\theta$, where A is the mass number of a fragment, $\theta$ its emission angle 
and P$_0$ the momentum per   $^8$B nucleon (P$_0$=2.0A GeV/c). Fig. \ref{fig:2} presents the P$_t$ distribution  for protons with 
the mean value $<$P$_T>$=73 MeV/c. Going over to the c.m.s. of $^7$Be+p compensates the transverse momentum transferred 
to the $^8$B nucleus which results in a far more narrow Pt distribution (insertion in Fig. \ref{fig:2}),  $<$P$_T^*>$= 52 MeV/c. 
Such a small value of P$_T$ is in agreement with data \cite{Smedberg99} in which for the longitudinal momentum 
distribution of fragments from $^8$B$\rightarrow^7$Be  reaction there was obtained a FWHM value of 91$\pm$5 MeV/c. 
In this way, the loose bound of the proton with the core is revealed under the most clear conditions of  observing 
fragmentation.
\par

\begin{figure}
\includegraphics[width=160mm]{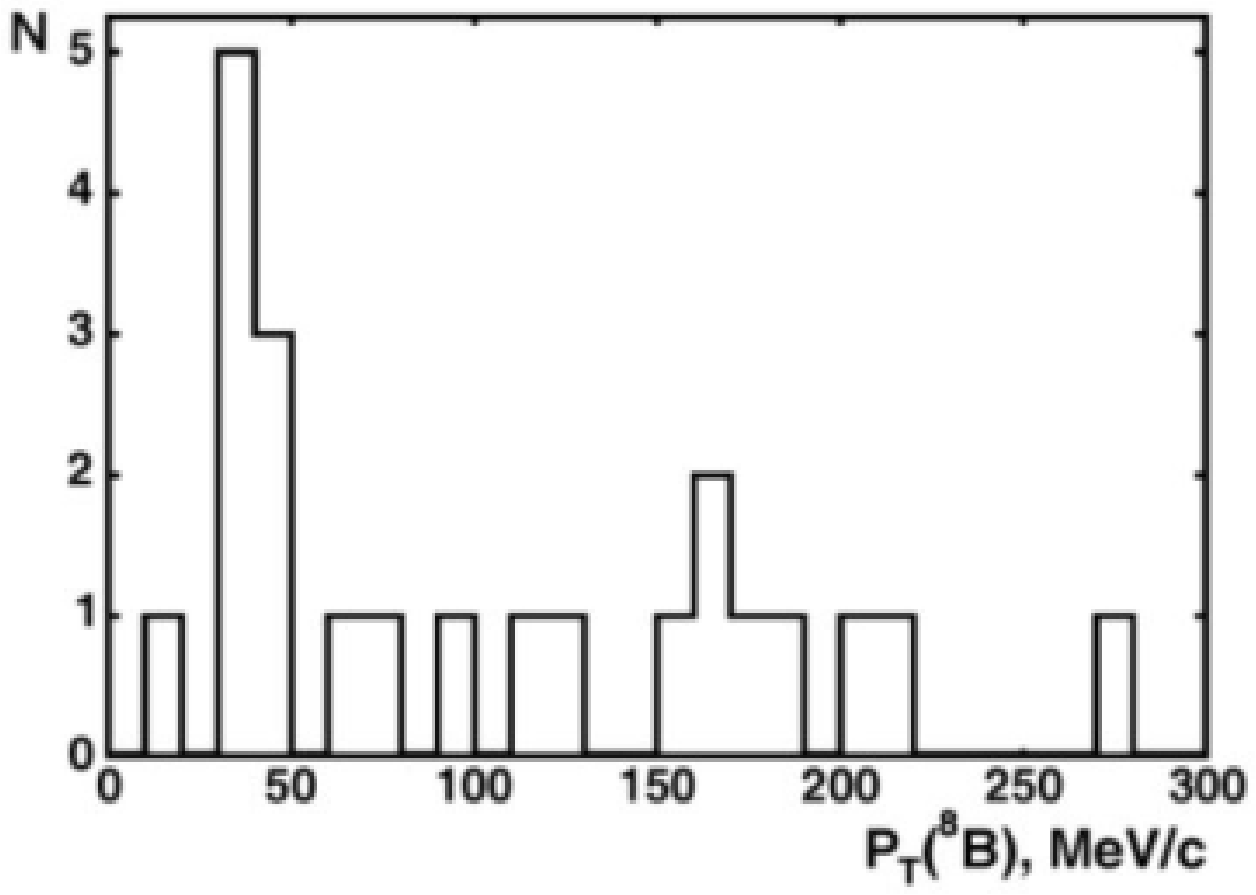} 
\caption{\label{fig:3} The P$_T$($^8$B) total transverse momentum distribution of $^7$Be+p pairs  produced 
in \lq\lq white\rq\rq ~stars $^8$B$\rightarrow^7$Be+p.}
\end{figure}

\indent Fig. \ref{fig:3} gives the distribution along the total  transverse momentum P$_T$($^8$B) which was acquired by  
the $^8$B nuclei  in the formation  of  \lq\lq white\rq\rq ~stars  $^7$Be+p; its mean value is about 100 MeV/c.
 The distribution peak is located at the P$_T$($^8$B) value of about 50 MeV/c. This asymmetry  of the P$_T$($^8$B) 
distribution  may be associated both with the emission of neutrons by the target-nuclei and the contribution from the 
core excitation  process $^8$B$\rightarrow^7$Be$^*$+p$\rightarrow^7$Be+$\gamma$+p \cite{Meister03}.
\par
\begin{figure}
\includegraphics[width=160mm]{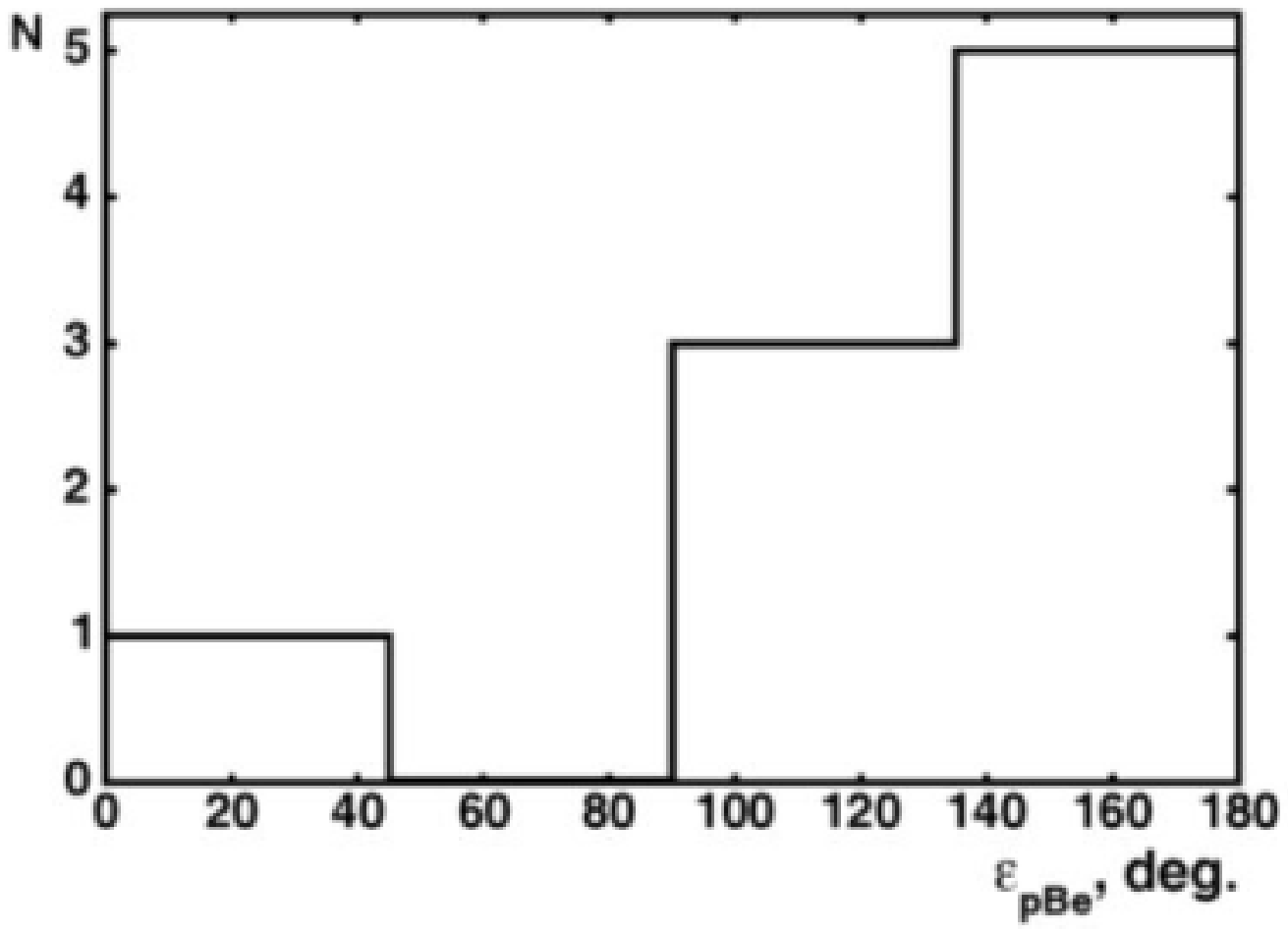} 
\caption{\label{fig:4} The $\epsilon_{pBe}$ azimuthal angle distribution in \lq\lq white\rq\rq ~stars 
$^8$B$\rightarrow^7$Be+p for P$_T$($^8$B)$<$60 MeV/c.}

\end{figure}
\indent It is possible to estimate  the role of the correlations by the azimuthal angle $\epsilon_{pBe}$ between $^7$Be 
and p. This distribution has an asymmetry  with respect to the angle  $\pi$/2 A$\approx$0.3, which points to a noticeable 
contribution of pairing dissociations. Fig. \ref{fig:4} presents the $\epsilon_{pBe}$ distribution with the requirement of selection
of the events P$_T$($^8$B)$<$60 MeV/c. One can note a sharp increase in the asymmetry up to A$\approx$0.7,
 which points to a binary character of the $^8$BB breakup  at small transverse momentum transfers.    	
\par
\section{\label{sec:level5}Conclusions}
\indent For the first time, nuclear emulsions were exposed to a beam of relativistic $^8$B nuclei. We have obtained data 
on the probabilities of  the $^8$B  fragmentation channels in peripheral interactions. 55 events of the peripheral 
$^8$B dissociation which do not involve the production of the target-nucleus fragments  and mesons 
(\lq\lq white\rq\rq ~stars ) were selected. A leading contribution of the $^8$B$\rightarrow^7$Be+p  mode having the lowest
 energy threshold was revealed on the basis of these events. Information about a relative probability of dissociation modes
 with larger multiplicity have been obtained. The  $^7$Be core dissociation in $^8$B is found to be similar to that of
 the free $^7$Be nucleus. A further analysis of the fragmentation topology suggests the identification of the H and He
 isotopes.
 \par
 \indent We have obtained the transverse momentum distributions for the $^8$B$\rightarrow^7$Be+p dissociation 
fragments. Their small average value, $<$P$_T^*>$=52 MeV/c, in the c.m.s. reflects a low binding energy of the external
 proton  in the $^8$B nucleus. In the selection of the events in which a transverse momentum of less than 60 MeV/c  is
 transferred to the $^8$B nucleus there appears a strong azimuthal angle correlation between $^7$Be and p. 
\par
\begin{acknowledgments}
\indent The work was supported by the Russian Foundation for Basic Research
 ( Grants 96-159623, 02-02-164-12a,03-02-16134, 03-02-17079 and 04-02-16593),
 VEGA 1/9036/02.  Grant from the Agency for Science of the Ministry for Education of the
 Slovak Republic and the Slovak Academy of Sciences, and Grants from the JINR
 Plenipotentiaries of the Republic of Bulgaria, the Slovak Republic, the Czech Republic
 and Romania in the years 2002-2005.\par
 \indent  The results presented here are based on a painstaking  job of the technicians-micro-scopists 
  I.I.Sosulnikova, A.M. Sosulnikova and G.V.Stelmakh  (JINR). A valuable contribution was made by the specialists 
who ensured a stable operation of the Nuclotron of the V.I.Veksler and A.M.Baldin Laboratory of High Energies. 
The authors are indebted to the Directorate of the G.N.Flerov Laboratory of Nuclear Reactions (JINR) for support in 
purchasing emulsions.     
\par  
\end{acknowledgments}

\end{document}